\documentclass[preprint, showpacs,preprintnumbers,amsmath,amssymb]{revtex4}

\usepackage{graphicx}
\usepackage{dcolumn}
\usepackage{bm}

\begin{document}

\preprint{APS/123-QED}

\title{Ion response in a weakly ionized plasma with ion flow}

\author{Roman Kompaneets}
\affiliation{School of Physics,
The University of Sydney, New South Wales 2006, Australia}

\author{Yuriy O. Tyshetskiy}
\affiliation{School of Physics, The University of Sydney, New South Wales 2006, Australia}

\author{Sergey V. Vladimirov}
\affiliation{School of Physics, The University of Sydney, New South Wales 2006, Australia}

\date{\today}

\begin{abstract}
We study the ion response to an initial perturbation in a weakly ionized plasma with 
ion flow driven by a dc electric field. 
The analysis is made by extending the classical Landau work [J. Phys. (USSR) {\bf 10}, 25 (1946)] 
to the ion kinetic equation including ion-neutral collisions and a dc electric field. 
We show, in particular, that the complex frequencies of ion waves
can be directly found from a known expression for
the ion susceptibility [Phys. Rev. E {\bf 71}, 016405 (2005); Phys. Rep. {\bf 27}, 997 (2001)];
this is not obvious from its original derivation, because it only aims to describe the 
ion response for real frequencies.

\end{abstract}

\pacs{52.30.-q, 52.25.Ya, 52.25.Mq, 52.25.Dg, 52.35.Fp, 52.35.Qz, 52.27.Lw}
\maketitle
\section{Introduction}
Low-pressure gas discharges are characterized by the presence of strong ion flow 
driven by the electric field that naturally arises to maintain the balance of 
absorption of ions and absorption of electrons on the electrodes and walls of the discharge chamber~\cite{fortov-khr-khr-phys.uspekhi-2004,
zeuner-mei-vacuum-1995,
nosenko-fis-mer-phys.plasmas-2007,
land-goe-new.j.phys-2006}. This field 
extends far beyond the pure ion sheath \cite{lieberman-1994, riemann-j.phys.d.appl.phys-2003},
so that the velocity of the ion flow driven by this field
often exceeds the thermal velocity of neutrals in the most part of the 
discharge (see, e.g., Fig.~7 of Ref.~\cite{land-goe-new.j.phys-2006}). 

This field-driven ion flow plays an important role in various phenomena in low-pressure gas discharges,
with the most obvious examples being ion waves~\cite{li-ma-li-phys.lett.a-2006,yoshimura-nak-wat-j.phys.soc.jap-1997,hershkowitz-ko-wan-ieee.trans.plasma.sci-2005}
and
interaction between charged dust particles levitated in discharges~\cite{ishihara-j.phys.d.appl.phys-2007,ishihara-vla-phys.plasmas-1997, vladimirov-ost-phys.rep-2004, 
morfill-ivl-rev.mod.phys-2009,
lampe-joy-gan-phys.plasmas-2000,
takahashi-ois-shi-phys.rev.e-1998, miloch-vla-ieee.trans.plasma.sci-2010, melzer-sch-pie-phys.rev.lett-1999}. 
To study the role of the ion flow in such phenomena, 
it is helpful to derive the ion susceptibility in a plasma in the presence
of field-driven ion flow.

The ion susceptibility in the presence
of field-driven ion flow has been self-consistently derived in 
Refs.~\cite{ivlev-zhd-khr-phys.rev.e-2005, schweigert-plasma.phys.rep-2001}. 
To clarify, the self-consistency here means two things. First, the steady-state velocity distribution is found from the model itself, i.e., from 
the balance of ion acceleration in the dc electric field driving the ion flow and ion-neutral collisions (instead of assuming a model distribution, e.g., a shifted Maxwellian distribution).
Second, collisions and the dc field
not only define the steady state but are also accounted for in the analysis of perturbations.
The resulting susceptibility is conveniently expressed in Ref.~\cite{ivlev-zhd-khr-phys.rev.e-2005}
via the plasma dispersion function [Eq.~(8) of Ref.~\cite{ivlev-zhd-khr-phys.rev.e-2005}].

The derivation of Refs.~\cite{ivlev-zhd-khr-phys.rev.e-2005, schweigert-plasma.phys.rep-2001}
aims to describe the ion response for real frequencies and real wave numbers and does not allow using
the derived susceptibility to find the complex frequencies of ion waves (for real wave numbers),
and the present paper addresses this issue.

Let us first explain why the derivation of Refs.~\cite{ivlev-zhd-khr-phys.rev.e-2005, schweigert-plasma.phys.rep-2001}
does not allow using the derived susceptibility to find the complex frequencies of ion waves.

We first note that the derivation of
Refs.~\cite{ivlev-zhd-khr-phys.rev.e-2005, schweigert-plasma.phys.rep-2001}
assumes the perturbations to be $\propto \exp(-i \omega t + i {\bf k} \cdot {\bf r})$, substitutes this to the ion kinetic equation, 
and finds the susceptibility as the normalized ratio of the complex amplitudes of the ion density and the potential.
[Of course, for real $\omega$ and ${\bf k}$ the perturbations of the above form can be considered as the Fourier components
of the actual perturbations (as the Fourier transform is defined for real $\omega$ and ${\bf k}$),
so that for real $\omega$ and ${\bf k}$ the susceptibility derived by the above method
has a clear meaning: it is the normalized proportionality coefficient between the 
Fourier components of the perturbations of
the ion density and the potential.]

We point out the following well-known fact:
in a collisionless one-component Maxwellian plasma in the absence of extraneous charges
there are {\it no solutions} where the perturbation of the distribution function and the 
perturbation of the potential are $\propto \exp(-i \omega t + i {\bf k} \cdot {\bf r})$ 
with the same complex $\omega$ and the same real ${\bf k}$ \cite{landau-j.phys.ussr-1946}.
(A one-component plasma is the
approximation where only one plasma component oscillates.)
Indeed, existing solutions in a collisionless one-component Maxwellian plasma
in the absence of extraneous charges
are characterized by different time dependencies
of the potential and the distribution function
(for the same ${\bf k}$): 
while oscillations of the potential are exponentially Landau damped, the perturbation of the distribution function 
experiences undamped oscillations due to free dispersal of particles ($\propto \exp[-i ({\bf k} \cdot {\bf v}) t]$ at large $t$, see Ref.~\cite{landau-j.phys.ussr-1946},
\S 1, last paragraph). Such solutions cannot be found when the perturbation of the potential and the perturbation of the distribution function
are assumed to be $\propto \exp(-i \omega t + i {\bf k} \cdot {\bf r})$ with the same $\omega$ and the same ${\bf k}$ (as assumed in the derivation of 
Refs.~\cite{ivlev-zhd-khr-phys.rev.e-2005, schweigert-plasma.phys.rep-2001}). Thus, it is not clear how the susceptibility
of Refs.~\cite{ivlev-zhd-khr-phys.rev.e-2005, schweigert-plasma.phys.rep-2001}) can be used to characterize such solutions.

Leaving aside the issue with the assumed form of the perturbations,
we also point out that
the validity of the mathematical calculations of the derivation of Refs.~\cite{ivlev-zhd-khr-phys.rev.e-2005, schweigert-plasma.phys.rep-2001}
is limited to the case where the perturbations do not decay faster than the ion-neutral collision frequency [i.e., where ${\rm Im}(\omega)$ is larger than minus the ion-neutral collision frequency],
as shown in Appendix~\ref{mistake}.

To find the complex frequencies of ion waves in the model of Refs.~\cite{ivlev-zhd-khr-phys.rev.e-2005, schweigert-plasma.phys.rep-2001},
one might suggest analytically continuing the susceptibility of Refs.~\cite{ivlev-zhd-khr-phys.rev.e-2005, schweigert-plasma.phys.rep-2001} 
to the lower half of the complex frequency plane, in analogy to a collisionless Maxwellian plasma \cite{landau-j.phys.ussr-1946},
but this requires substantiation. Indeed, the correctness of this procedure for a collisionless Maxwellian plasma 
has been justified by solving the initial value problem \cite{landau-j.phys.ussr-1946},
and it is not known how the solution of the initial value problem 
for the model of Refs.~\cite{ivlev-zhd-khr-phys.rev.e-2005, schweigert-plasma.phys.rep-2001}
looks like. In particular, it is not clear whether
the same method (as in Ref.~\cite{landau-j.phys.ussr-1946}) of calculation of the integral restoring the potential from its Laplace transform
in time will work for the model of Refs.~\cite{ivlev-zhd-khr-phys.rev.e-2005, schweigert-plasma.phys.rep-2001}.

To address the issue, 
we solve the initial value problem
for the model of Refs.~\cite{ivlev-zhd-khr-phys.rev.e-2005, schweigert-plasma.phys.rep-2001}.
That is, we consider an arbitrary initial perturbation and analyze its time evolution.
This allows us to confirm the above hypothesis that
the complex frequencies of ion waves
can be found by analytically continuing the susceptibility of Refs.~\cite{ivlev-zhd-khr-phys.rev.e-2005, schweigert-plasma.phys.rep-2001} 
to the lower half of the complex frequency plane.
Since in Refs.~\cite{ivlev-zhd-khr-phys.rev.e-2005, schweigert-plasma.phys.rep-2001} the derived susceptibility is presented in such a form
that its analytical continuation to the lower half of the complex frequency plane is given by the same expression,
it can be said that the 
complex frequencies of ion waves
can be found from the same expression for the susceptibility as presented in Refs.~\cite{ivlev-zhd-khr-phys.rev.e-2005, schweigert-plasma.phys.rep-2001} 
by simply considering this expression as a function of complex $\omega$.

The susceptibility of Refs.~\cite{ivlev-zhd-khr-phys.rev.e-2005, schweigert-plasma.phys.rep-2001}
has already been used in Ref.~\cite{kompaneets-pre} to study ion waves,
and thus our work substantiates the validity
of the study reported in Ref.~\cite{kompaneets-pre}. We further note that, as our derivation can be easily generalized to a 
multiple-ion-species plasma, our work is of importance to the potential use
of the susceptibility of Refs.~\cite{ivlev-zhd-khr-phys.rev.e-2005, schweigert-plasma.phys.rep-2001}
to study ion waves in the presence of multiple ion species.

\section{Model}
\label{model}
We consider a weakly ionized plasma in a dc electric field ${\bf E}_0$ driving ion flow. 
We assume that electrons obey a Boltzmann distribution with a very large temperature,
so we consider their number density $n_0$ to be homogeneous and fixed
(note that Boltzmann electron distributions, in the presence of field-driven ion flow,
are common in low-pressure gas discharges \cite{riemann-j.phys.d.appl.phys-2003, hershkowitz-ko-wan-ieee.trans.plasma.sci-2005}; 
we address the role of the electron temperature
in Sec.~\ref{discussion}).
We assume ${\bf E}_0$ to be homogeneous and use
the kinetic equation for ions 
with the Bhatnagar-Gross-Krook  (BGK) ion-neutral collision term \cite{bhatnagar-gro-kro-phys.rev-1954, ivlev-zhd-khr-phys.rev.e-2005,tolias-rat-ang-phys.plasmas-2011}
and Poisson's equation:
\begin{eqnarray}
\frac{\partial f}{\partial t}
+{\bf v} \cdot \frac{\partial f}{\partial {\bf r}}
+\frac{e}{m} \left( {\bf E}_0 - \frac{\partial \phi}{\partial {\bf r}} \right)
\cdot \frac{\partial f}{\partial {\bf v}} \nonumber \\
 = -\nu f +\nu \Phi_{\rm M} \int f({\bf v}') \, d{\bf v}',
\label{kinetic-equation}
\end{eqnarray}
\begin{equation}
-\bigtriangleup \phi = \frac{e}{\epsilon_0} \left( \int f \, d{\bf v} -n_0 \right),
\label{poisson}
\end{equation}
where $f$ is the ion velocity distribution function,
$\phi$ is the electric potential
describing the time-space varying field
(i.e., the field apart from ${\bf E}_0$),
\begin{equation}
\Phi_{\rm M}=\frac{1}{(2\pi v_{\rm tn}^2)^{3/2}}\exp\left(-\frac{v^2}{2v_{\rm tn}^2}\right)
\label{neutral-distribution}
\end{equation}
is the normalized Maxwellian velocity distribution of neutrals,
$\nu$ is the ion-neutral collision frequency, which is assumed to be velocity-independent,
$v_{\rm tn}=\sqrt{k_{\rm B}T_{\rm n}/m}$ is the thermal velocity of neutrals,
$T_{\rm n}$ is the temperature of neutrals,
$e$ is the elementary charge (ions are assumed to be singly ionized),
$m$ is the ion mass, $k_{\rm B}$ is the Boltzmann constant, and $\epsilon_0$ is the permittivity of free space.
Note that the BGK term exactly 
describes charge transfer collisions under the assumption
of a velocity-independent collision frequency, as explained in Ref.~\cite{ivlev-zhd-khr-phys.rev.e-2005}.

The homogeneous steady-state solution $f=f_0$ is found from
Eqs.~(\ref{kinetic-equation}) and (\ref{poisson}) by setting $\phi=0$, 
$\partial f/\partial t =0$, $\partial f/\partial {\bf r}={\bf 0}$.
The resulting velocity distribution $f_0$ is not a shifted Maxwellian distribution and can be written as \cite{ivlev-zhd-khr-phys.rev.e-2005}
\begin{equation}
f_0= \frac{n_0} {(2\pi \upsilon_{\rm tn}^2)^{3/2}}\int_0^{\infty}\exp\left(-\xi -\frac{|{\bf v}-\xi{\bf v}_{\rm f}|^2}{2\upsilon_{\rm tn}^2}\right) \, d{\xi},
\label{distribution}
\end{equation}
where 
\begin{equation}
{\bf v}_{\rm f}= \frac{e{\bf E}_0}{m\nu}
\end{equation}
and the subscript ``f'' stands for ``flow'', as the flow velocity $(1/n_0)\int {\bf v} f_0 \, d{\bf v}$ can be shown to be equal to
${\bf v}_{\rm f}$.
Equation~(\ref{distribution}) here is Eq. (3) of
Ref.~\cite{ivlev-zhd-khr-phys.rev.e-2005}, but rewritten using another integration variable
in order to show that $f_0$ is an integral superposition of shifted Maxwellian distributions with exponential
weights.

We consider a small arbitrary initial perturbation of the ion distribution function
and determine the resulting evolution of the potential $\phi$.
It is sufficient to consider the perturbations to be of the form
\begin{eqnarray}
f({\bf r}, {\bf v}, t)-f_0({\bf v})=f_1({\bf v}, t) \exp(i {\bf k} \cdot {\bf r}), \nonumber \\
\phi({\bf r}, t)=\phi_1(t) \exp(i {\bf k} \cdot {\bf r}),
\end{eqnarray}
(with a real ${\bf k}$), since in the general case one can expand the perturbations into a Fourier integral in space
and then consider each Fourier component separately. We denote $f_1({\bf v}, 0)$ by $f_{1, {\rm i}}({\bf v})$
(where the subscript ``i'' stands for ``initial'') and determine $\phi_1(t)$.

\section{Result}
\label{result-sec}
In this section we formulate the result, while its derivation is provided in the next section. 
We first note that there are two cases, namely whether the initial perturbation satisfies certain reasonable
conditions (stated in Sec.~\ref{derivation}) or not. These conditions are satisfied for a wide class of functions $f_{1, {\rm i}}({\bf v})$;
an example of an initial perturbation satisfying these conditions is $f_{1,{\rm i}}({\bf v}) \propto \exp[-v^2/(2w^2)]$,
where $w$ is an arbitrary real constant. 

If the initial perturbation satisfies
these conditions, then the evolution of the potential is described by:
\begin{eqnarray}
\phi_1(t)= \sum_{{\rm Im}(\omega_j)>\Gamma} C_j \exp (-i \omega_j t)
+ H( t, \Gamma)\exp(\Gamma t),
\label{solution}
\end{eqnarray}
where $C_j$ are certain coefficients [for their explicit form, see Eq.~(\ref{c-coefficients})], the complex frequencies $\omega_j$ are the roots
of the dispersion relation
\begin{equation}
1+\chi(\omega)=0,
\label{dispersion-relation}
\end{equation}
the index $j$ numbers these roots
(there is an infinite number of these roots, even when $E_0=0$ and $\nu \to 0$,
and these roots are known as the Landau plasma modes \cite{derfler-sim-phys.fluids-1969}),
the sum is only over those $j$ that satisfy ${\rm Im}(\omega_j)>\Gamma$,
$\Gamma$ is an arbitrary real number
(e.g., it can be chosen to be highly negative so that the $\exp(\Gamma t)$ factor in the second term
of the right-hand side of Eq.~(\ref{solution}) decays with time very rapidly),
the function $\chi(\omega)$ is given by the same expression as the susceptibility derived in Refs.~\cite{ivlev-zhd-khr-phys.rev.e-2005, schweigert-plasma.phys.rep-2001}:
\begin{subequations}
\label{chi}
\begin{eqnarray}
\chi(\omega)=\frac{\omega_{\rm pi}^2}{\nu^2}
\frac{B(\omega)}{1-A(\omega)}, 
\label{chi-1} \\
A(\omega)=\int_0^\infty \exp [-\Psi(\omega, \eta)] \,d{\eta}, \label{chi-2}  
\label{chi-a}\\
B(\omega)=\int_0^\infty 
\frac{
\eta \exp [-\Psi(\omega, \eta)]} 
{1+i({\bf k} \cdot {\bf v}_{\rm f}/\nu)\eta }
\,d{\eta}, \label{chi-3} \label{chi-b} \\
\Psi(\omega, \eta)=
\left(1-\frac{i\omega}{\nu}\right)\eta \nonumber \\
+\frac{1}{2}
\left[
\frac{i{\bf k} \cdot {\bf v}_{\rm f}}{\nu} 
+\left(\frac{kv_{\rm tn}}{\nu}\right)^2
\right]
\eta^2, \label{chi-4} 
\end{eqnarray}
\end{subequations}
$H(t, \Gamma)$ is a certain function [for its explicit form, see Eq.~(\ref{h-function})] and tends to zero as $t \to \infty$
(as shown in the last but one paragraph of Sec.~\ref{derivation}), i.e., at large $t$ the second term of the right-hand side of Eq.~(\ref{solution})
is negligible as compared to any non-zero term of the sum represented by the first term of the right-hand side of Eq.~(\ref{solution}), and
$\omega_{\rm pi}=[n_0 e^2/(\epsilon_0 m)]^{1/2}$ is the ion plasma frequency. Equation~(\ref{solution}) shows that
the asymptotic behavior of $\phi_1(t)$ at large $t$ is independent of the properties of the initial perturbation,
as the frequencies $\omega_j$ are determined entirely by the plasma parameters and the wave number ${\bf k}$. The frequencies $\omega_j$ are analyzed in Ref.~\cite{kompaneets-pre}.

If the initial perturbation does not satisfy the conditions mentioned in the beginning of this section, then 
the asymptotic behavior of $\phi_1(t)$ at large $t$
may be determined not only by the plasma parameters and the wave number but also by the
properties of the initial perturbation, in contrast to the case considered above.

Note that we also provide an
expression for the potential $\phi_1(t)$ valid irrespectively of which of the above two cases holds; this expression is Eq.~(\ref{phi-final-simple}).

\section{Derivation}
\label{derivation}
To derive the result stated above, we need to linearize Eqs.~(\ref{kinetic-equation}) and (\ref{poisson}) with respect to
$f_1$ and $\phi_1$ and then solve the resulting linearized equations with respect to $\phi_1$. The linearization gives:
\begin{eqnarray}
\frac{\partial f_1}{\partial t}
+i {\bf k}\cdot {\bf v} f_1
+\frac{e{\bf E}_0}{m} \cdot \frac{\partial f_1}{\partial {\bf v}}
-\frac{i{\bf k}e\phi_1}{m} \cdot
\frac{\partial f_0}{\partial {\bf v}}= \nonumber \\
 = -\nu f_1 +\nu \Phi_{\rm M} \int f_1({\bf v}') \, d{\bf v}',
\label{kinetic-equation-linearized}
\end{eqnarray}
\begin{equation}
k^2 \phi_1 = \frac{e}{\epsilon_0} \int f_1 \, d{\bf v}.
\label{poisson-linearized}
\end{equation}

We solve Eqs.~(\ref{kinetic-equation-linearized}) and (\ref{poisson-linearized}) by using the Laplace transform in time. The Laplace transform in time 
is defined here as
\begin{equation}
\hat \psi (\omega)= \int_0^{\infty} \psi(t) \exp (i \omega t) \, dt,
\label{transforms}
\end{equation}
where 
the hat denotes the application of the transform and $\psi(t)$ is an arbitrary function. The inverse transform restoring
$\psi(t)$ from $\hat \psi (\omega)$ is:
\begin{eqnarray}
\psi(t) = \frac{1}{2\pi} \int_{-\infty+i\gamma}^{+\infty+i\gamma} \hat \psi (\omega) \exp (-i \omega t) \, d\omega,
\label{inverse-transforms}
\end{eqnarray}
where the integration is performed in the complex plane over the line ${\rm Im} (\omega) = \gamma$, and $\gamma$
is any real number such that this line is in the region of convergence of the integral in Eq.~(\ref{transforms}).
Taking the Laplace transform of Eqs.~(\ref{kinetic-equation-linearized}) and
(\ref{poisson-linearized}), i.e., multiplying these equations by $\exp (i \omega t)$ and then integrating over $\int_0^{\infty}\, dt$,
we get
\begin{eqnarray}
(-i \omega + \nu + i {\bf k}\cdot {\bf v}) \hat f_1 +\frac{e {\bf E}_0}{m} \cdot
\frac{\partial \hat f_1}{\partial {\bf v}} = \frac{i {\bf k} e \hat \phi_1}{m} \cdot \frac{\partial f_0}{\partial {\bf v}} \nonumber \\
+ \nu \Phi_{\rm M} \int \hat f_1({\bf v}')\, d{\bf v}' + f_{1,{\rm i}},
\label{kinetic-equation-transformed}
\end{eqnarray}
\begin{eqnarray}
k^2 \hat \phi_1 = \frac{e}{\epsilon_0} \int \hat f_1 \,d{\bf v}.
\label{poisson-equation-transformed}
\end{eqnarray}
Thus the initial value problem is reduced to solving 
Eqs.~(\ref{kinetic-equation-transformed}) and (\ref{poisson-equation-transformed}) with respect to $\hat \phi_1$.

Let us reduce solving Eqs.~(\ref{kinetic-equation-transformed}) and (\ref{poisson-equation-transformed}) with respect to $\hat \phi_1$
to finding the solution of Eq.~(\ref{kinetic-equation-transformed}) with respect to $\hat f_1$ for a fixed $\hat \phi_1$.
To do so, we first note that the latter solution has the following structure: 
\begin{equation}
\hat f_1 = L_1 \hat \phi_1 + L_2 \int \hat f_1({\bf v}') \, d{\bf v}' + {\cal L}(f_{1,{\rm i}}),
\label{kinetic-equation-transformed-solution}
\end{equation}
where the first, second and third terms in the right-hand side denote the direct contributions from the first, second and third terms of the right-hand 
side of Eq.~(\ref{kinetic-equation-transformed}), respectively,
$L_{1,2}$ are certain coefficients (which depend on ${\bf v}$), ${\cal L}$ is a certain linear operator, 
and we do not need explicit expressions for $L_{1,2}$ and ${\cal L}$ at the moment.
[Note that the solution (\ref{kinetic-equation-transformed-solution}) is unique because of
the boundary condition $\left. \hat f_1 \right|_{{\bf v} \to \infty}=0$.]
We integrate Eq.~(\ref{kinetic-equation-transformed-solution}) over ${\bf v}$,
which allows us to express the integral $\int \hat f_1 \, d{\bf v}$ via $L_{1,2}$, ${\cal L}$, and $\hat \phi_1$. By substituting the result to Eq.~(\ref{poisson-equation-transformed}) and using 
Eq.~(\ref{inverse-transforms}), we get
\begin{eqnarray}
\phi_1(t) = \frac{1}{2\pi} \int_{-\infty+i\gamma}^{+\infty + i\gamma} d\omega \, \exp(-i\omega t)
\nonumber \\
\times
\frac{          
(e/\epsilon_0)  \int {\cal L}    (f_{1,{\rm i}})     \, d{\bf v}\           
}
{             
 k^2[1-\int L_2 \, d{\bf v}]         -         (e/\epsilon_0)  \int L_1 \, d{\bf v}          
 }.
\label{phi-final}
\end{eqnarray}
Thus, to solve the initial value problem, 
there only remains to derive explicit expressions for $L_{1,2}$ and ${\cal L}$.

Let us now solve Eq.~(\ref{kinetic-equation-transformed}) with respect to $\hat f_1$ for a fixed $\hat \phi_1$
in order to derive the required expressions for $L_{1,2}$ and ${\cal L}$.
Making the substitution 
\begin{equation}
\hat f_1 = \beta \exp (-\alpha), 
\end{equation}
where 
\begin{equation}
\alpha=\frac{m}{eE_0} \left[ (-i \omega + i k_x v_x + \nu)v_z + \frac{1}{2} i k_z v_z^2 \right],
\label{alpha-definition}
\end{equation}
the $z$-axis is in the direction of ${\bf E}_0$, and the $x$-axis is directed so that $k_y=0$, we reduce Eq.~(\ref{kinetic-equation-transformed})
to
\begin{equation}
\frac{\partial \beta}{\partial v_z} = \frac{m \exp(\alpha)}{eE_0}[\ldots],
\label{kinetic-equation-transformed-ready-to-integrate}
\end{equation}
where $[\ldots]$ denotes the right-hand side of Eq.~(\ref{kinetic-equation-transformed}). 
To integrate Eq.~(\ref{kinetic-equation-transformed-ready-to-integrate}), we need a boundary condition on $\beta$,
so let us find a boundary condition on $\beta$ from that $\hat f_1=0$ at $v_z \to \pm \infty$.
Let us explicitly write the condition that $\hat f_1=0$ at $v_z \to \pm \infty$ in terms of $\beta$: $\beta \exp (-\alpha) = 0$ at $v_z \to \pm \infty$.
The behavior of $\exp (-\alpha)$ at $v_z \to \pm \infty$ is determined by the sign of ${\rm Im}(\omega)+\nu$ [as seen from Eq.~(\ref{alpha-definition})], but
we only need to 
consider the case
\begin{equation}
{\rm Im}(\omega)+\nu>0
\label{limitation}
\end{equation}
because in Eq.~(\ref{phi-final}) we can choose any $\gamma$
larger than a certain threshold.
Considering this case, we note that 
$\exp(-\alpha)$ does not tend to zero as $v_z \to -\infty$ and tends to zero as $v_z \to +\infty$, as seen from Eq.~(\ref{alpha-definition}).
Thus, we obtain that $\beta=0$ at $v_z \to -\infty$; note that one cannot set $\beta=0$ at $v_z \to +\infty$.
Using that $\beta=0$ at $v_z \to -\infty$, we integrate Eq.~(\ref{kinetic-equation-transformed-ready-to-integrate})
to obtain
\begin{eqnarray}
L_1=\frac{i}{E_0} 
\int_{-\infty}^{v_z}
\, dv_z' \, \exp(\alpha'-\alpha) 
\left[ k_z \frac{\partial f_0'}{\partial v_z'}-\frac{v_x k_x}{v_{\rm tn}^2}f_0'\right],
\label{l1}
\end{eqnarray}
\begin{equation}
L_2=\frac{1}{v_{\rm f}} \int_{-\infty}^{v_z} \, dv_z' \, \exp(\alpha '-\alpha) \Phi_{\rm M}',
\end{equation}
\begin{equation}
{\cal L}(f_{1,{\rm i}})=\frac{m}{eE_0} \int_{-\infty}^{v_z} \, dv_z' \, \exp(\alpha'-\alpha) f_{1,{\rm i}}',
\label{l-operator}
\end{equation}
where all quantities denoted by the prime symbol contain $v_z'$ instead of $v_z$.

Thus we have in principle solved the initial value problem, as the solution
is provided by Eqs.~(\ref{phi-final}) and (\ref{l1})-(\ref{l-operator}), with $\gamma$ being a sufficiently large number
such that the result of the integration in Eq.~(\ref{phi-final}) for any larger $\gamma$ is the same.

Let us obtain a simpler form for the potential by simplifying the integrals $\int L_{1,2} \, d{\bf v}$ 
and $\int {\cal L}(f_{1,i})\, d{\bf v}$ present in Eq.~(\ref{phi-final}).

Let us start with the integrals $\int L_{1,2} \, d{\bf v}$. We first perform the integration over $v_y$
and then over $v_x$ using the formula
\begin{equation}
\int_{-\infty}^{+\infty} x^{1,2} \exp \left( -\frac{x^2}{2}+c x\right) \, d{x} =  i c^{1,2} \sqrt{2\pi} \exp \left( -\frac{c^2}{2}\right)
\label{integration-formula}
\end{equation}
valid for complex $c$. The next step is to replace $v_z'$ by $v_z-v_{\rm f}\eta$, where $\eta$ is a new integration variable, and to integrate over $v_z$ using 
Eq.~(\ref{integration-formula}). Then the integration over $\xi$, the variable originating from Eq.~(\ref{distribution}), is elementary. These calculations yield 
$\int L_{1} \, d{\bf v}=-(k^2 n_0eB)/(m\nu^2)$ and $\int L_{2} \, d{\bf v}=A$, where $A$ and $B$ are defined by Eqs.~(\ref{chi-a}) and (\ref{chi-b}), respectively. 
Thus, the denominator in Eq.~(\ref{phi-final}) can be written as $(1+\chi)k^2$, where $\chi$ is given by Eq.~(\ref{chi}).

Let us now simplify the integral $\int {\cal L}( f_{1,{\rm i}})\, d{\bf v}$. 
We first change the order of integration, 
$\int_{-\infty}^{+\infty}\, d{v_z} \int_{-\infty}^{v_z} \, d{v_z'} \to \int_{-\infty}^{+\infty}\, d{v_z'} \int_{v_z'}^{\infty} \, d{v_z}$.
Then we replace $v_z$ by $v_z'+q$, where $q$ is a new integration variable,
and integrate over $q$ using the formula
\begin{eqnarray}
\int_0^{\infty}\exp\left(-ax - \frac{1}{2} i b x^2 \right) \, d{x} = \frac{1-i\, {\rm sign}(b)}{2}\sqrt{\frac{\pi}{|b|}} \nonumber \\
\times \exp\left[ \left(
\frac{[i\,{\rm sign}(b)-1]a}{2\sqrt{|b|}}
\right)^2\right]
\nonumber \\
\times
\left[
1+{\rm erf}
\left(
\frac{[i\,{\rm sign}(b)-1]a}{2\sqrt{|b|}}
\right)
\right]
\end{eqnarray}
valid for real $a>0$ and real $b \not = 0$,
where 
\begin{equation}
{\rm erf}(z)=\frac{2}{\sqrt{\pi}} \int_0^z \exp(-\xi^2)\, d\xi
\end{equation}
is the error function
defined for a complex variable.
After that we can remove the prime symbol in $v_z'$.
The result can be written as:
\begin{eqnarray}
\int {\cal L}(f_{1, {\rm i}})\, d{\bf v}=
-\frac{{\rm sign}(k_z)+i}{2}\sqrt{\frac{m}{eE_0 |k_z|}}
\int f_{1,{\rm i}} Z(\Lambda) \, d{\bf v},
\end{eqnarray}
where 
\begin{eqnarray}
Z(x)= 2i \exp(-x^2)\int_{-\infty}^{i\xi} \exp(-\xi^2) \, d\xi
\label{plasma-dispersion-function}
\end{eqnarray}
is the plasma dispersion function
and
\begin{eqnarray}
\Lambda=\frac{{\rm sign}(k_z)+i}{2}
\sqrt{\frac{m}{eE_0 |k_z|}}
\nonumber \\
\times (-i \omega + \nu + i {\bf k}\cdot {\bf v}).
\label{large-lambda}
\end{eqnarray}

Using the above obtained results for $\int L_{1,2} \, d{\bf v}$ and $\int {\cal L}(f_{1,{\rm i}})\, d{\bf v}$,
we write the potential (\ref{phi-final}) as
\begin{eqnarray}
\phi_1(t) = -\frac{{\rm sign}(k_z)+i}{4\pi \epsilon_0 k^2}\sqrt{\frac{me}{E_0 |k_z|}}
\nonumber \\
\times
\int_{-\infty + i\gamma}^{+\infty + i \gamma} \, d\omega \, \exp(-i \omega t)
\frac{\int f_{1, {\rm i}} Z (\Lambda) \, d{\bf v}}{1+\chi},
\label{phi-final-simple}
\end{eqnarray}
where $\chi$ is given by Eq.~(\ref{chi}); 
again, we note that $\gamma$ is a sufficiently large number
such that the result of the integration in Eq.~(\ref{phi-final-simple}) for any larger $\gamma$ is the same.

\begin{figure}
\includegraphics[width=7cm]{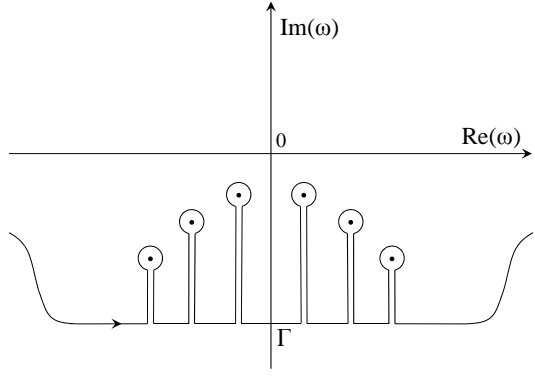}
\caption{Contour of integration used to derive Eq.~(\ref{solution}). The dots denote the solutions of the dispersion relation (\ref{dispersion-relation}).}
\label{shema}
\end{figure}

Let us now calculate the integral in Eq.~(\ref{phi-final-simple}) 
by using Cauchy's integral theorem and assuming certain reasonable conditions on the initial perturbation.
To do so, we consider 
the fraction in the last line of Eq.~(\ref{phi-final-simple}) 
as a function of $\omega$ in the entire complex $\omega$ plane
[despite that we only obtained the expressions for $L_{1,2}$ and ${\cal L}$
for the region (\ref{limitation})]
and modify the contour of integration over $\omega$ in such a way that (i) Cauchy's integral theorem guarantees that 
the value of the integral is not changed and (ii) integration over the new contour yields Eq.~(\ref{solution}).

We assume that the initial perturbations satisfies the following conditions: (i) the expression in the numerator in the last line of Eq.~(\ref{phi-final-simple}) 
defines an analytic function of $\omega$ 
in the entire complex $\omega$ plane
and (ii) this function is sufficiently well-behaved at ${\rm Re}(\omega)\to \pm \infty$
so that there are no problems with the validity of the results stated below.
We verified numerically that conditions (i) and (ii) above are met 
for a wide class of functions $f_{1,{\rm i}}({\bf v})$
[e.g., for $f_{1,{\rm i}}({\bf v}) \propto \exp(-v^2/2w^2)$,
where $w$ is an arbitrary real constant].
We displace the integration contour down as shown in Fig.~\ref{shema}.
This, according to Cauchy's integral theorem,
does not change the value of the integral, as the 
function $1/(1+\chi)$ [where $\chi$ is defined by Eq.~(\ref{chi}) in the entire complex $\omega$ plane] 
is analytic in the entire complex $\omega$ plane except at the zeros
of the function $1+\chi$. The integral over the new contour is the sum
of the contribution of the circles shown in Fig.~\ref{shema}
and the contribution of the horizontal part. This allows us to write Eq.~(\ref{solution}),
where
\begin{eqnarray}
C_j=\frac{1-i\, {\rm sign}(k_z)}{2 \epsilon_0 k^2}\sqrt{\frac{me}{E_0 |k_z|}}
\left. \frac{\int f_{1, {\rm i}} Z (\Lambda) \, d{\bf v}}{d\chi/d\omega} \right|_{\omega=\omega_j},
\label{c-coefficients}
\end{eqnarray}
\begin{eqnarray}
H(t, \Gamma) = 
-\frac{{\rm sign}(k_z)+i}{4\pi \epsilon_0 k^2}\sqrt{\frac{me}{E_0 |k_z|}} \nonumber \\
\times \int_{-\infty}^{\infty} d\omega_{\rm r} \, \exp(-i \omega_{\rm r} t) \left. \left( \frac{\int f_{1, {\rm i}} Z (\Lambda) \, d{\bf v}}{1+\chi} \right) \right|_{\omega=\omega_{\rm r}+ i \Gamma},
\label{h-function}
\end{eqnarray}
and the number $\Gamma$ defines the position
of the horizontal part of the new contour as shown in Fig.~\ref{shema};
we emphasize that in Eqs.~(\ref{c-coefficients}) and (\ref{h-function}),
$\chi$ and $\Lambda$ are given by
Eqs.~(\ref{chi}) and (\ref{large-lambda})
{\it in the entire complex $\omega$ plane}
despite that we only obtained the expressions for $L_{1,2}$ and ${\cal L}$
for the region (\ref{limitation});
the first term of the right-hand side of Eq.~(\ref{solution}) is the contribution of the circles shown in Fig.~\ref{shema},
and the second term of the right-hand side of Eq.~(\ref{solution}) is the contribution of the horizontal part;
we introduced a new integration variable $\omega_{\rm r}={\rm Re}(\omega)$ (where the subscript ``r'' stands for ``real'')
to conveniently write the contribution of the horizontal part;
the integration over $\omega_{\rm r}$ in Eq.~(\ref{h-function}) is performed over the real axis.

To substantiate our claim made in Sec.~\ref{result-sec} that the function $H(t, \Gamma)$
tends to zero as $t \to \infty$,
we note that in the right-hand side of Eq.~(\ref{h-function}), $t$ is only present in $\exp(-i \omega_{\rm r} t)$, so that the integral in Eq.~(\ref{h-function}) can be considered as the inverse Fourier transform in time.
Hence, this integral tends to zero as $t \to \infty$, according to the Riemann-Lebesgue lemma
[indeed, the numerator in this integral is a sufficiently well-behaved function, as assumed in the beginning of the previous paragraph;
the denominator in this integral tends to $1$ as $\omega_{\rm r} \to \pm \infty$, as can be easily verified].

Let us now briefly discuss the case where the initial perturbation does not satisfy the conditions stated in the beginning of the second paragraph back.
To deal with this case, one can analytically continue the numerator
in the last line of Eq.~(\ref{phi-final-simple}) into the
region below the line ${\rm Im}(\omega)=\gamma$. To do so, one may need to specify
cuts; the obtained analytical continuation may also have singular points.
Then one can attempt to 
modify the procedure
detailed in the second paragraph back by modifying the contour shown in Fig.~\ref{shema};
in particular, the contour should now also pass around the aforementioned cuts and/or singularities so that
Cauchy's integral theorem can be applied.
The contributions from the contour parts that pass around the aforementioned cuts and/or singularities
should be added to Eq.~(\ref{solution}). The time dependence of these additional terms will be determined
not only by the plasma parameters but also by the properties of the initial perturbation.

\section{Role of the electron temperature}
\label{discussion}
Let us discuss the role of the electron temperature.
There are two effects related to a finite electron temperature: (i) the electron response to ion oscillations,
and (ii) a finite inhomogeneity length of the Boltzmann electron distribution in the field ${\bf E}_0$.
Concerning effect (i), it can be taken into account by
adding the term
$1/(k \lambda_{\rm e})^2$ to the left-hand side of Eq.~(\ref{dispersion-relation}), where $\lambda_{\rm e}=[\epsilon_0 k_{\rm B} T_{\rm e}/(n_0 e^2)]^{1/2}$ is the electron Debye length
and $T_{\rm e}$ is the electron temperature.
As regards effect (ii), the corresponding inhomogeneity 
length is $L_{\rm e} = k_{\rm B}T_{\rm e}/(eE_0)$, and
our model applies when this distance is larger than both 
the ion-neutral collision length, which is ${\rm max} \{v_{\rm f}, v_{\rm tn}\}/\nu$,
and the wavelength $2\pi/k$.
This means the following applicability limitation of the model:
\begin{equation}
\frac{k_{\rm B} T_{\rm e}}{m} \gg {\rm max} \left\{ v_{\rm f}^2, \, v_{\rm tn} v_{\rm f}, \, \frac{v_{\rm f}\nu}{k} \right\}.
\label{limitation1}
\end{equation}

\section{Conclusion}
We have demonstrated that the expression for the ion susceptibility derived in 
Refs.~\cite{ivlev-zhd-khr-phys.rev.e-2005, schweigert-plasma.phys.rep-2001}
[and given by Eq.~(\ref{chi}) of our paper]
can be used in the entire complex frequency plane 
to find the complex frequencies of ion waves, despite that
its original derivation~\cite{ivlev-zhd-khr-phys.rev.e-2005, schweigert-plasma.phys.rep-2001}
only aims to describe the ion response for real frequencies.

\begin{acknowledgments}
The authors thank Alexei Ivlev for useful discussions.
R.~K. acknowledges
the receipt of a
Professor Harry Messel Research Fellowship supported by
the Science Foundation for Physics within the University of
Sydney.
The work was partially supported by the Australian Research Council. 
\end{acknowledgments}

\appendix
\section{Note on the derivation of Refs.~\cite{ivlev-zhd-khr-phys.rev.e-2005, schweigert-plasma.phys.rep-2001}}
\label{mistake}
In this Appendix we substantiate our claim made in Introduction that 
the validity of the mathematical calculations of the derivation of Refs.~\cite{ivlev-zhd-khr-phys.rev.e-2005, schweigert-plasma.phys.rep-2001}
is limited to the case ${\rm Im}(\omega)+\nu>0$.
To do so, we repeat the derivation of Refs.~\cite{ivlev-zhd-khr-phys.rev.e-2005, schweigert-plasma.phys.rep-2001}
and point out where this restriction arises.
In this Appendix we use the notations of the present paper and not the notations of Refs.~\cite{ivlev-zhd-khr-phys.rev.e-2005, schweigert-plasma.phys.rep-2001};
the model of our paper and the model of Refs.~\cite{ivlev-zhd-khr-phys.rev.e-2005, schweigert-plasma.phys.rep-2001} are the same. 

The idea of the derivation of Refs.~\cite{ivlev-zhd-khr-phys.rev.e-2005, schweigert-plasma.phys.rep-2001} is to assume that the perturbations are
of the form
\begin{eqnarray}
f({\bf r}, {\bf v}, t)-f_0({\bf v})=f_{\rm a}({\bf v}) \exp(-i \omega t + i {\bf k} \cdot {\bf r}), \\
\phi({\bf r}, t)=\phi_{\rm a} \exp(-i \omega t + i {\bf k} \cdot {\bf r}),
\end{eqnarray}
(here the subscript ``a'' stands for ``amplitude''),
substitute this to the kinetic equation linearized with respect to the perturbations, solve 
this equation with respect to $f_{\rm a}$ for a fixed $\phi_{\rm a}$, and find the susceptibility as
\begin{equation}
\chi=-\frac{e}{\epsilon_0k^2}\frac{\int f_{\rm a} \, d{\bf v}}{\phi_{\rm a}}.
\label{susceptibility-definition}
\end{equation} 
Concerning the first step, the linearized kinetic equation takes the form:
\begin{eqnarray}
(-i \omega + \nu + i {\bf k}\cdot {\bf v}) f_{\rm a} +\frac{e {\bf E}_0}{m} \cdot
\frac{\partial f_{\rm a}}{\partial {\bf v}} = \frac{e i {\bf k \phi_{\rm a}}}{m} \cdot \frac{\partial f_0}{\partial {\bf v}} \nonumber \\
+ \nu \Phi_{\rm M} \int f_{\rm a}({\bf v}')\, d{\bf v}'.
\label{kinetic-equation-linearized-schweigert}
\end{eqnarray}

To solve the linearized kinetic equation [Eq.~(\ref{kinetic-equation-linearized-schweigert})] with respect to $f_{\rm a}$,
one makes the substitution 
\begin{equation}
f_{\rm a}=F \exp(-\alpha)
\label{substitution-appendix}
\end{equation}
where $\alpha$ is defined by Eq.~(\ref{alpha-definition}). The resulting equation is
\begin{equation}
\frac{\partial F}{\partial v_z} = \frac{m \exp(\alpha)}{eE_0}[\ldots],
\label{kinetic-equation-transformed-ready-to-integrate-schweigert}
\end{equation}
where [$\ldots$] denotes the right-hand side of Eq.~(\ref{kinetic-equation-linearized-schweigert}).

To integrate the resulting equation [Eq.~(\ref{kinetic-equation-transformed-ready-to-integrate-schweigert})], one needs a boundary condition
on $F$, and here there are two different cases, namely whether ${\rm Im}(\omega)+\nu$ is positive or negative, as explained in the following.
The boundary condition on $f_{\rm a}$ is
\begin{equation}
f_{\rm a}=0 \quad {\rm at} \quad v_z \to \pm \infty.
\label{boundary1}
\end{equation}
Let us explicitly write this condition in terms of $F$ using Eq.~(\ref{substitution-appendix}):
\begin{equation}
F \exp(-\alpha)=0 \quad {\rm at} \quad v_z \to \pm \infty.
\label{boundary2}
\end{equation}
The behavior of $\exp(-\alpha)$ at $v_z \to \pm \infty$ depends on the sign of ${\rm Im}(\omega)+\nu$, as seen from the definition of $\alpha$ [Eq.~(\ref{alpha-definition})], 
so there are two different cases:
\begin{enumerate}
\item ${\rm Im}(\omega)+\nu>0$. In this case $\exp(-\alpha)$ does not tend to zero as $v_z \to -\infty$ and tends to zero as $v_z \to +\infty$, as seen from Eq.~(\ref{alpha-definition}). 
Using this fact and Eq.~(\ref{boundary2}),
we get the following boundary condition on $F$: 
\begin{equation}
F = 0 \quad {\rm at} \quad  v_z \to - \infty
\label{boundary-case1}
\end{equation}
(i.e., one cannot set $F=0$ for $v_z \to +\infty$). Having obtained Eq.~(\ref{boundary-case1}), we do not need to know the behavior of $F$
at $v_z \to +\infty$ in order to integrate Eq.~(\ref{kinetic-equation-transformed-ready-to-integrate-schweigert}), 
as 
Eq.~(\ref{kinetic-equation-transformed-ready-to-integrate-schweigert}) is 
a first-order differential equation.
\item ${\rm Im}(\omega)+\nu<0$. In this case, in contrast to the preceding case, 
$\exp(-\alpha)$ tends to zero as $v_z \to -\infty$ and not as $v_z \to +\infty$, as seen from Eq.~(\ref{alpha-definition}).
Thus we obtain the following boundary condition on $F$:
\begin{equation}
F = 0 \quad {\rm at} \quad v_z \to +\infty
\label{boundary-case2}
\end{equation}
(i.e., one cannot set $F=0$ for $v_z \to -\infty$).
\end{enumerate}

The derivation of Refs.~\cite{ivlev-zhd-khr-phys.rev.e-2005, schweigert-plasma.phys.rep-2001} only considers case 1 above
and is thus limited to the case ${\rm Im}(\omega)+\nu>0$. 
Using Eq.~(\ref{boundary-case1}),
one integrates Eq.~(\ref{kinetic-equation-transformed-ready-to-integrate-schweigert}) to obtain:
\begin{eqnarray}
f_{\rm a}= \int_{-\infty}^{v_z} dv_z \, \exp(\alpha'-\alpha)
\left[ 
\frac{i\phi_{\rm a}}{E_0}\left( 
k_z \frac{\partial f_0'}{\partial v_z'}
\right.
\right.
\nonumber \\
\left.
\left.
-\frac{v_x k_x}{v_{\rm tn}^2}f_0'
\right) +\frac{\Phi_{\rm M}'}{v_{\rm f}}
\int f_{\rm a}({\bf v}'') \, d{\bf v}''
\right],
\label{schweigert-integration-result}
\end{eqnarray}
where all quantities denoted by the prime symbol contain $v_z'$ instead of $v_z$, and the $x$-axis
is directed so that $k_y=0$.
Equation~(\ref{schweigert-integration-result}) above is Eq.~(5) of Ref.~\cite{schweigert-plasma.phys.rep-2001}
[one needs to clarify here that Eq.~(5) of Ref.~\cite{schweigert-plasma.phys.rep-2001} is incorrectly typed,
as the quantities inside the large round brackets in Eq.~(5) of Ref.~\cite{schweigert-plasma.phys.rep-2001} should contain $v_z'$ instead of $v_z$].

Let us demonstrate that Eq.~(\ref{schweigert-integration-result}) does not apply in case 2 above.
To demonstrate this, in the following we show that in case 2 Eq.~(\ref{schweigert-integration-result}) does not satisfy the boundary condition $f_{\rm a} =0$ at $v_z \to +\infty$. 
We consider the particular case $k_z=0$ (for simplicity)
and rewrite Eq.~(\ref{schweigert-integration-result}) so as to 
explicitly show the dependence of $f_{\rm a}$ on $v_z$ at $v_x=0$:
\begin{eqnarray}
f_{\rm a} \propto \exp\left[ -\frac{mv_z}{eE_0} \left( -i \omega + \nu \right)\right]
\nonumber \\
\times \int_{-\infty}^{v_z} dv_z' \, \exp\left[ \frac{mv_z'}{eE_0} 
\left( -i \omega + \nu \right) 
-\frac{v_z'^2}{2 v_{\rm tn}^2}
\right].
\label{dependence-on-vz}
\end{eqnarray}
As $v_z \to +\infty$, neither the integral in Eq.~(\ref{dependence-on-vz}) nor the exponent
in front of this integral tends to zero, so we see that the condition $f_{\rm a} = 0$ at $v_z \to +\infty$ is indeed violated.

\end{document}